\documentclass[8pt,a4paper]{article}
\usepackage[superscript]{cite}
\usepackage{amsmath}
\usepackage{graphicx}
\usepackage{caption}
\usepackage{hyperref}
\usepackage{geometry}
\usepackage{authblk}
\usepackage{abstract}
\usepackage{siunitx}
\usepackage{array}
\usepackage{multirow}
\usepackage{hhline}
\usepackage{enumitem}
\usepackage{tablefootnote}
\usepackage[flushleft]{threeparttable}
\usepackage{textcomp}
\title{Inverted Pyramid 3-axis Silicon Hall Effect Magnetic Sensor With Offset Cancellation}

\author[1]{Jacopo Ruggeri}
\author[2]{Udo Ausserlechner}
\author[2]{Helmut Köck}
\author[1 *]{Karen M. Dowling}
\affil[1]{Department of Microelectronics, Delft University of Technology, Delft, The Netherlands}
\affil[2]{Infineon Technologies AG, 9500 Villach, Austria}

\date{}

\begin{document}

\maketitle

\vspace{-3em} % Adjust this value to reduce space between title and the next section

% Extra section for Corresponding and Contributing Authors
\begin{center}
    *\textbf{Corresponding author:} K.M.Dowling@tudelft.nl\\
    \textbf{Contributing authors:} J.Ruggeri@tudelft.nl, Udo.Ausserlechner@infineon.com, Helmut.Koeck@infineon.com
\end{center}

\vspace{0.2em}

\begin{abstract}
Microelectronic magnetic sensors are essential in diverse applications, including automotive, industrial, and consumer electronics. Hall-effect devices hold the largest share of the magnetic sensor market, and they are particularly valued for their reliability, low cost and CMOS compatibility.
This paper introduces a novel 3-axis Hall-effect sensor element based on an inverted pyramid structure, realized by leveraging MEMS micromachining and CMOS processing. The devices are manufactured by etching the pyramid openings with TMAH and implanting the sloped walls with n-dopants to define the active area. Through the use of various bias-sense detection modes, the device is able to detect both in-plane and out-of-plane magnetic fields within a single compact structure. In addition, the offset can be significantly reduced by one to three orders of magnitude by employing the current-spinning method. The device presented in this work demonstrated high in-plane and out-of-plane current- and voltage-related sensitivities ranging between \SI{64.1}{} to \SI{198}{\volt\per\ampere\per\tesla} and \SI{14.8}{} to \SI{21.4}{\milli\volt\per\volt\per\tesla}, with crosstalk below \SI{3.7}{\percent}. The sensor exhibits a thermal noise floor which corresponds to approximately \SI{0.5}{\micro\tesla\per\sqrt{\hertz}} at \SI{1.31}{\volt} supply.
This novel Hall-effect sensor represents a promising and simpler alternative to existing state-of-the-art 3-axis magnetic sensors, offering a viable solution for precise and reliable magnetic field sensing in various applications such as position feedback and power monitoring.
\end{abstract}

\section*{Introduction}
Magnetic sensors serve as fundamental components in a wide array of applications, from industrial and automotive systems to consumer electronics and medical devices. 
The ability to accurately sense three-dimensional magnetic fields is increasingly crucial, enabling precise position sensing, enhanced motion tracking for robotic and commercial products, contactless angle measurement in automotive steering systems and biomagnetic measurements. \cite{application1,application2,applications3}. %Depending on the application, several types of magnetic sensors are commonly adopted, each with its own set of advantages and disadvantages \cite{sensor_types,sensor_types2}.
As technological advancements drive the demand for miniaturization and integration in electronic devices, microscopic and integrated magnetic sensors have become particularly important. The most prevalent types of microelectronic magnetic sensors are magnetoresistive (xMR), MEMS and Hall-effect sensors.\cite{sensor_types3} 
Despite the recent progress in MEMS and xMR devices, Hall-effect sensors still hold the largest share of the magnetic sensor market thanks to their durability, reliability, and cost-effectiveness. Compared to their counterparts, Hall-effect devices display lower sensitivity and SNR, but they are very easily integrated in a CMOS process, making them a viable and inexpensive solution suitable for commercial applications \cite{Hall_effect_cit}. 

\begin{figure}[t!]
  \centering
  \includegraphics[width=5in]{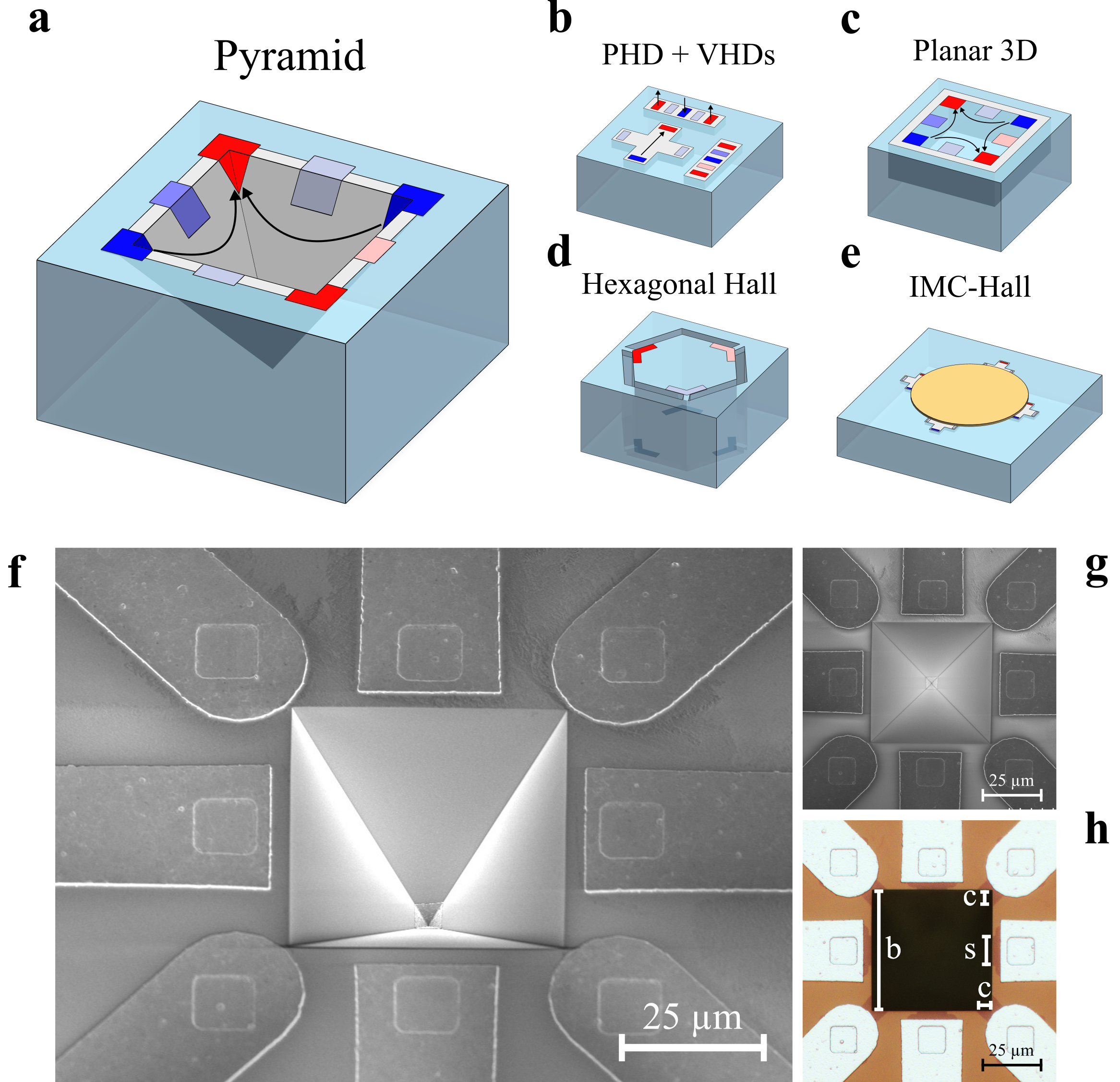}
  \caption{\textbf{State-of-the-art and 3-axis pyramid device.} \textbf{a} Our novel pyramid Hall device. \textbf{b} A standard 3-axis Hall sensor, composed of one planar Hall device and two vertical Hall devices. \textbf{c} The flat 8-contact Hall device, proposed by Schott.\cite{SCHOTT2000167} \textbf{d} Sander's hexagonal prism Hall device.\cite{SANDER2016587} \textbf{e} An IMC Hall device.\cite{popovic2001integrated} \textbf{f} SEM image of the pyramid device, with a tilting angle of \SI{30}{\degree}. \textbf{g} SEM image of the pyramid device, with no tilting angle. \textbf{h} Optical microscope image of the sensor. The pyramid size and the contact dimensions are highlighted in the picture.}
  \label{fig:1}
\end{figure}

The two main types of Hall effect sensors are vertical Hall devices (VHD) and planar Hall devices (PHD). PHDs detect the out-of-plane component of the field, while VHDs detect the in-plane components of the field. When integrated on the same substrate (\autoref{fig:1}b) they provide a cheap and simple solution for 3D magnetic field sensing \cite{Pascal}. However, the performance of this configuration is limited by the vertical Hall devices. In fact, VHDs require the supply current to flow deep into the active area to reach reasonable sensitivities\cite{VHD_popovich}. This is achieved by means of deep, low-doped wells that, however, are not available in many CMOS technologies. This issue can be solved by scaling down the tub dimensions of the vertical element, but the very close spacing of the contacts causes the device to operate in velocity saturation, which increases the non-ideal effects such as residual offset and flicker noise.\cite{Ausserlechner}

To overcome these issues, as well as to reduce the number of terminals and to reduce footprint, other concepts of 3-axis Hall effect devices have been proposed.
For instance, Schott et al. \cite{SCHOTT2000167} proposed a planar, eight-contact 3-axis Hall effect sensor (\autoref{fig:1}c). This device combines a bottomless Hall plate with coupled 3-contact VHDs to be compact and have high sensitivity. However, similarly to VHDs, it requires deep active areas to operate well. In addition, no current spinning technique is described, so the low offset and low noise of this device are strictly related to the adopted buried technology, which further limits its integration into a CMOS process. 
Sanders et al. \cite{Sander_table,SANDER2016587} designed and manufactured a six-contact hexagonal Hall device (\autoref{fig:1}d), with high sensitivity and isotropy. In addition, the device can be current-spinned to modulate offset and flicker noise. However, the device has a large footprint and the fabrication is quite complex since it involves double-side patterning, implantation, and deep reactive ion etching.

Lastly, Popovich's approach to the problem was to deposit, in a post-processing step, a ferromagnetic material called integrated magneto-concentrator (IMC) on top of several Hall plates \cite{popovic2001integrated}. An example of IMC sensor is represented in \autoref{fig:1}e.  The ferromagnetic material converts locally the in-plane magnetic field into an out-of-plane magnetic field, which can be detected by combining the output of several PHDs. This configuration presents the high sensitivity and low offset typical of Hall plates, and it can be integrated in a CMOS process with few post-processing steps. However, these devices may exhibit hysteresis issues, higher temperature drifts, and saturation at high magnetic fields due to the use of ferromagnetic materials. In addition, this sensor is inherently anisotropic.\cite{schott_IMC}

In this work, we present a novel type of 3-axis Hall effect sensor, based on an inverted pyramid structure. The device, represented in \autoref{fig:1}a, can detect the three components of the magnetic field within a single, compact structure using different biasing and sensing configurations, also called modes. Previously, we demonstrated the basic device concept by employing three different modes to perform simple magnetic field measurements \cite{Ruggeri}. Now, we additionally prove that each of these modes can be current-spun, which means that the offset can be reduced by one to three orders of magnitude. This is a notable result, since offset calibration is a time consuming, costly process in industry which additionally requires zero background magnetic field in the production line. Current-spinning removes this need, improving production efficiency. In addition, we characterize multiple first-generation devices to extract the current-spun sensitivity and residual offset, as well as the crosstalk and the noise power spectrum. %With this work, we aim to demonstrate that this device is a valid alternative to existing solutions and has the potential to pave the way for a new class of integrated 3-axis magnetic sensors.

\begin{figure}[t!]
  \centering
  \includegraphics[width=\textwidth]{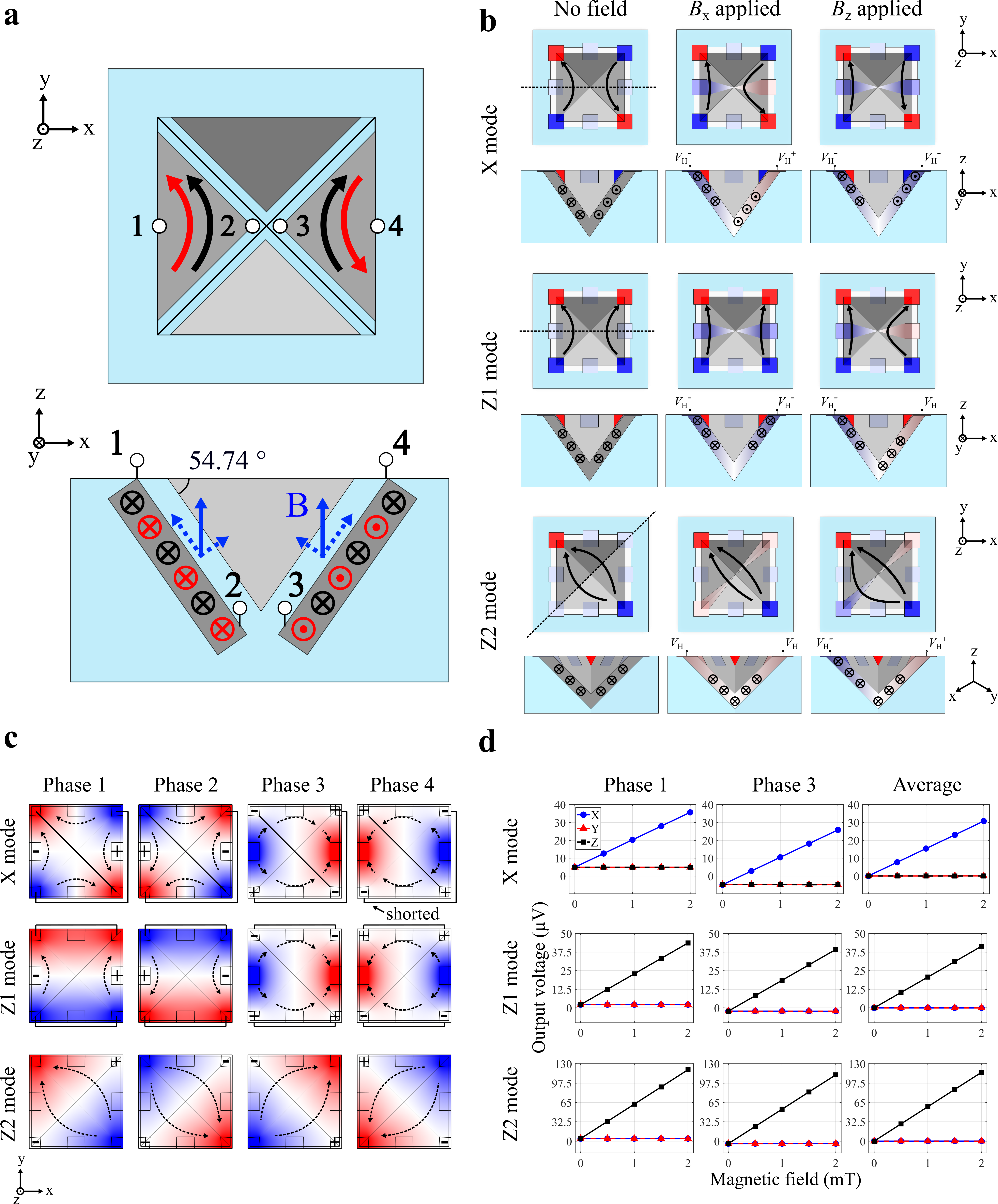}
  \caption{\textbf{Sensing mechanism of the 3-axis pyramid device.} \textbf{a} Simplified model of the pyramid device. The sensor is treated as four separated Hall plates. The black arrows represent the biasing configuration for z-field detection, while the red arrows display the biasing scheme for x-field sensing.  \textbf{b} Three sensing modes: X, Z1 and Z2. The black arrows represent the carrier velocities, and their deflection displays qualitatively the effect of an applied magnetic field. \textbf{c} The spinning sequence of modes X, Z1 and Z2. In mode X and Z1, the opposite and adiacent corner contacts are respectively shorted. The black arrows represent again the electrons' velocities. \textbf{d} COMSOL simulations of the spinning sequences. Two orthogonal phases are simulated and averaged per mode, to reproduce current-spinning. Both sensitivity and crosstalk are reported.}
  \label{fig:2}
\end{figure}
\section*{Sensor concept}

\subsection*{Sensor microstructure}
SEM and optical microscope images of the sensor are displayed in \autoref{fig:1}f-h.
The manufactured and characterized device is an eight-contact inverted pyramid with a characteristic angle of \SI{54.74}{\degree} and a square base \textit{b} of \SI{50}{\micro\meter}. Four contacts are positioned at the corner of the base, while the remaining four are located at the midpoint of each side. The active area is defined by doping the sloped sidewalls, resulting in a pyramid-shell structure. The contact locations and dimensions are defined by n+ implantations, which partially overlap with the active area to ensure electrical connectivity and ohmic contacts. The side wall contacts (s) are \SI{10}{\micro\meter} wide and the corner contacts (c) are \SI{5}{\micro\meter}. These heavily doped regions extend outside the active area to facilitate the connection with the metal traces, avoiding patterning issues related to the sloped walls. The entire device is covered by a passivation layer of silicon oxide, providing protection and stability to the underlying structures. The contact between the metal traces and the n+ regions is performed through contact openings or vias in the passivation layer, ensuring reliable electrical connections. The metal traces extend away from the pyramid structure and terminate with contact pads for further packaging. 
\subsection*{Working principle}
To clearly explain the working principle of the device, let us consider an approximate model with three key assumptions:
\begin{enumerate}[noitemsep]
  \item The four sloped faces of the inverted pyramid are separate Hall elements, each slanted at an angle of \SI{54.74}{\degree}.
  \item The four devices are identical, and no mismatch due to processing is present.
  \item Each device is ideal with no doping inhomogeneities or lithography inaccuracies/errors. 
\end{enumerate}
In this analysis, let us consider the two devices that are symmetric through the yz plane, as depicted in \autoref{fig:2}a. To begin, let us push two parallel and identical currents in the Hall sensors.
Under the listed assumptions, the Hall voltage produced by a single element between its middle side contact (1) and the apex of the pyramid (2) in the presence of an arbitrary magnetic field can be written as:
\begin{equation}
    V_H^{12} = Ssin(\theta)B_{x} + Scos(\theta)B_{z}
    \label{Hall_12}
\end{equation}
where $\theta = \SI{54.74}{\degree}$ is the etching angle and S, instead, is the sensitivity of a standard Hall plate:
\begin{equation}
    S = \frac{1}{2}G_H\frac{r_H}{ned}I_{supply} = \frac{1}{2}\mu_H\frac{G_H}{(\frac{L}{W})_{eff}}V_{supply}
    \label{current_sens}
\end{equation}
In the equation, $G_H$ is the geometric correction factor, $r_H$ is the Hall factor, $n$ is the concentration of carriers, e is the elementary charge, $d$ is the thickness of the plate, $\mu_H = r_H\mu $ is the Hall mobility and $(L/W)_{eff}$ is the effective number of squares.
The extra $\frac{1}{2}$ factor, not present in a typical Hall plate, is due to the fact the total supply current $I_{supply}$ is split between two elements.

If we consider the opposite device, the Hall voltage measured between the apex of the pyramid (3) and its middle side contact (4) can instead be written as:
\begin{equation}
    V_H^{34} = -S'sin(\theta)B_{x} + S'cos(\theta)B_{z}
    \label{Hall_23}
\end{equation}
where $S = S'$ because the device structures are assumed identical. \autoref{Hall_12} and \autoref{Hall_23} are a set of linear equations from which the x- and z- components of the field can be extracted.
The two equations can be added together to obtain:
\begin{equation}
    V_H^z = V_H^{12} + V_H^{34} = 2Scos(\theta)B_z  
\end{equation}
The Hall voltage $V_H^z$ is, therefore, only sensitive to the out-of-plane component of the field, and the crosstalk to the in-plane component is discarded by symmetry.
To obtain the in-plane sensitivity we can subtract \autoref{Hall_12} and \autoref{Hall_23}, or reverse the direction of the current that flow in the left device. In this way, the measured Hall voltage can be written as:
\begin{equation}
    V_H^{x} = V_H^{12} - V_H^{34} = 2Ssin(\theta)B_{x}  
\end{equation}
As already mentioned, the previous analysis is only valid under the assumption of perfectly identical devices. If this assumption is removed, we obtain: 
\begin{equation}
    V_H^z = (S+S')cos(\theta)B_z + (S-S')sin(\theta)B_{x}
\end{equation}
\begin{equation}
    V_H^{x} = (S-S')cos(\theta)B_z + (S+S')sin(\theta)B_{x}
\end{equation}
From this simple model, it can be understood that the crosstalk arises from asymmetries between two different sloped faces, such as doping inhomogeneities, lithography inaccuracies and contact misplacement.
In a more general situation where each device might present some imperfections, we may write:
\begin{equation}
    V_H^z =  S_{zx}''B_{x} + S_{zy}''B_{y} + S_{zz}''B_{z}
\end{equation}
\begin{equation}
    V_H^{x} = S_{xx}''B_{x} + S_{xy}''B_{y} + S_{xz}''B_{z}
\end{equation}
where the $S''_{nm}$ terms belong to the sensitivity matrix that relate the two Hall voltage components to the three components of the magnetic field. The analytical determination of the $S''$ coefficients is out of the scope of this work. %but it can be reconstructed from symmetry considerations, the tilting angle of the current and \autoref{current_sens}.
In this case, an additional crosstalk component is present from the other lateral dimension (y).
The crosstalks to the y-field are due to inaccuracies within the same device, which slightly tilt the currents. Since these currents do not flow anymore exclusively along the y-axis, a small 
Hall voltage is produced by the y-component of the magnetic field. 

Lastly, a practical pyramid device is more complex since current also flows across the device edges, which were previously assumed isolated. Even for a perfectly symmetric and uniform sensor, the sensitivity $S$ of a single sloped face presents $G_H$ and $(L/W)_{eff}$ factors that differ from the isolated assumption. In the pyramid device, in fact, the current can split between adjacent faces and re-enter the same element at a different location. This implies that different modes might display different geometry responses, even if they can be obtained just by swapping one out of two currents. What is more, now the apex contact is same (points (2) and (3) in \autoref{fig:2}a), and due to superposition we can simply measure the total voltage between the two sidewall contacts (points (1) and (2)). Therefore, the apex does not require an external electrical contact.
\subsection*{Operation modes}
For this study, we present three different operation modes named X/Y, Z1, and Z2, represented in \autoref{fig:2}b. In the X/Y mode, an alternating biasing pattern is applied, resulting in the flow of four anti-parallel currents. Two middle-side contacts, aligned along the x-axis, detect the Hall voltage variation due to an x-field. The other two middle-side contacts, instead, detect the y-field. The sensing mechanism is as previously explained: the two anti-parallel currents respond to the in-plane field with equal but opposite Lorentz forces. The electrons that flow in the first sloped side move closer to the sensing contact, while the electrons in the second face are dragged towards the apex of the pyramid. In the presence of a z-field, the electrons on the two faces are dragged by the same Lorentz force in the same direction, and the crosstalk is removed by symmetry.
In the Z1 mode, the bias pattern is symmetric with respect to the yz plane. Two parallel currents flow in the devices along the y-direction. The working principle is the same as in the X/Y mode, but the sensitivity and crosstalk components are exchanged.
The Z2 mode presents only one pair of supply contacts, located at opposite corners. The Hall voltage is sensed at the remaining corner contacts. Once the supply is applied, the current splits into two branches that flow parallel to each other, similar to the Z1 mode.

\subsection*{Current-spinning schemes}

In a real device, the output voltage ($V_{out}$) can be written as:
\begin{equation}
V_{out} = V_{H} + V_{off} + v_{n}(t)    
\end{equation}
where $V_H$ is the Hall voltage, $V_{off}$ is the offset voltage and $v_{n}$ is the noise of the system.
The offset represents the output signal at zero applied magnetic field and arises due to asymmetries provoked by mismatches, lithographic inaccuracies, and doping inhomogeneities. Typically, the offset is similar or even higher than the signal itself. Various methods have been developed to mitigate the offset, and the most commonly adopted is a dynamic offset cancellation technique known as current spinning.
Current spinning involves continuously switching the supply and sensing contacts over time. This process modulates the offset at higher frequencies while keeping the Hall voltage constant \cite{MUNTER1990743}. Despite this technique, some offset remains, known as the residual offset. In typical Hall devices, the offset reduction achieved is 500-1000 times the initial raw offset \cite{SANDER201692}.

The pyramid device, which features eight contacts, can be read out by a substantial number of biasing and sensing configurations. The objective is to identify orthogonal phases among these configurations to effectively cancel out the offset and inverse polarity phases to counteract thermal gradients. Moreover, these configurations must be sensitive only to the specific component of the magnetic field we aim to detect, with minimal crosstalk. For each mode proposed in the previous paragraph, the corresponding spinning sequence was identified and they are illustrated in \autoref{fig:2}c.
The first sequence represented is the X mode spinning scheme. The Y mode is not shown since it is identical to the X mode, but \SI{90}{\degree} rotated. 
Opposite corner contacts must be shorted to properly spin the device. As a consequence of this short, the device becomes a six-contact configuration. However, since we use only four contacts for either X or Y detection, it can be considered a four-contact device, similar to a standard planar Hall device. This simplification makes the spinning sequence straightforward. Notably, phases 1 and 2, and phases 3 and 4 are inverse, achieved by reversing the sign of the supply and sense contacts. Phases 1 and 3, as well as phases 2 and 4, are orthogonal and obtained by swapping the supply and sense terminals.
For the Z1 spinning sequence, the approach is similar to mode X/Y, but in this case, adjacent pairs of corner electrodes are shorted. Both modes X/Y and Z1 exhibit two orthogonal phases that are not symmetric since supply and sense contacts are of two different types, i.e. either corners or middle-side terminals.
Lastly, the Z2 spinning sequence does not require any short, it uses only the four corner contacts and the spinning sequence is exactly the same as that of a standard planar Hall device \cite{dowling_spinning}.
To prove the concept and functionality of each spinning sequence, two orthogonal phases were simulated using the finite element method (\autoref{fig:2}d). The simulation results indicate that, for a symmetric device, each phase of each mode does not exhibit crosstalk. Additionally, the offset related to the mesh is perfectly canceled by averaging the results of phases 1 and 3, since the simulation model assumes perfect electrical linearity. 
\newpage
\begin{figure}[h!]
  \centering
  \includegraphics[width=\textwidth]{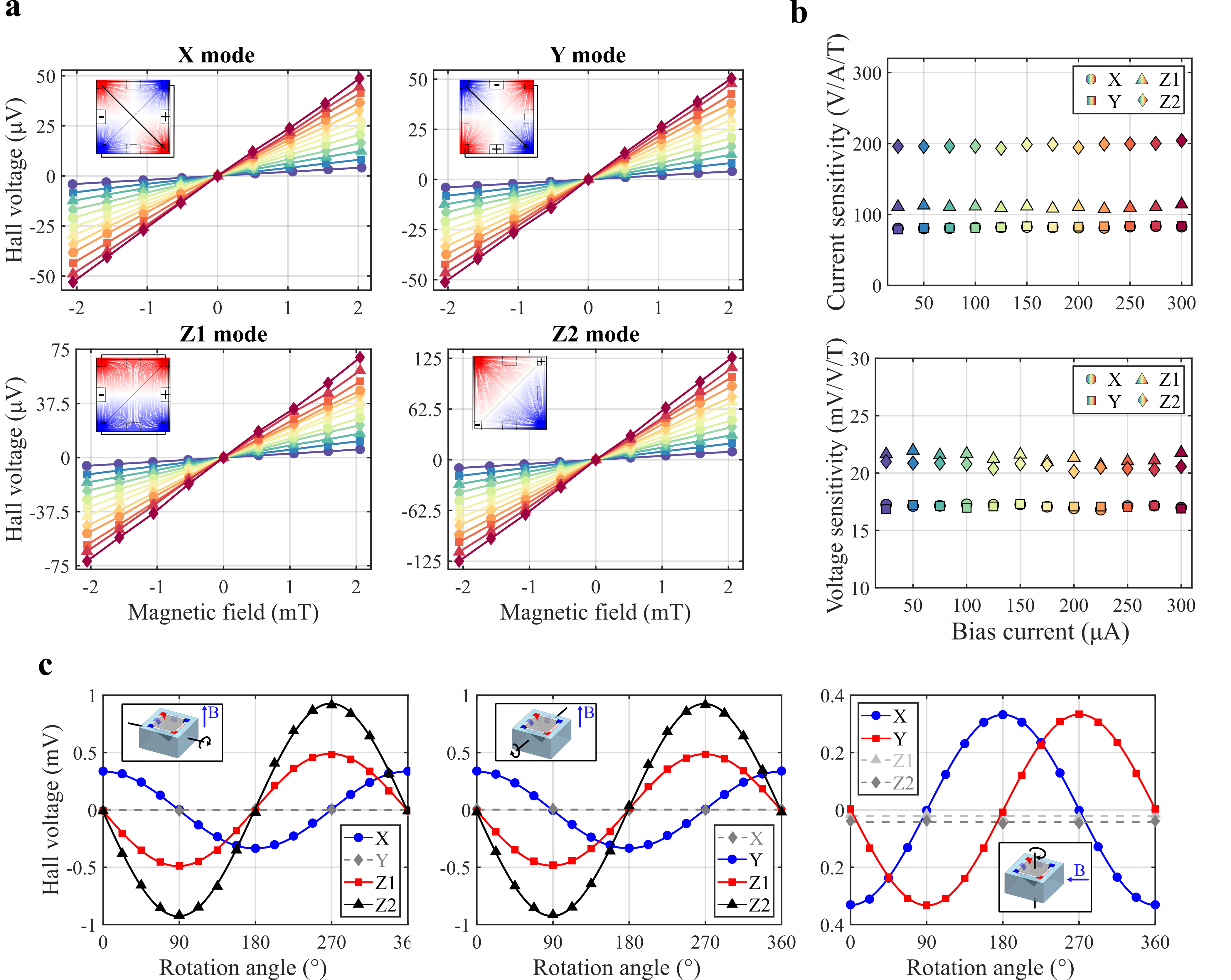}
  \caption{\textbf{Sensitivity and crosstalk measurements.} \textbf{a} Hall voltage in function of the magnetic field, for twelve equidistant supply currents in the range \SIrange{25}{300}{\micro\ampere}, measured for sensor P1. \textbf{b} Current- and voltage-related sensitivities of sensor P1, for twelve different supply currents. \textbf{c} Hall voltage in function of the rotation angle, for a \SI{100}{\micro\ampere} supply current and a magnetic field of \SI{48}{\milli\tesla}, measured for sensor P4. The solid line is the fit employed to extract the crosstalk. The rotation angle and direction of the magnetic field are reported in the inset of each graph.}
  \label{fig:3}
\end{figure}
\section*{Results}
\subsection*{Sensitivity and Crosstalk}

To evaluate the performance of each detection mode, the sensitivity to magnetic field was extracted when supplied with twelve equidistant supply currents in the range \SIrange[]{25}{300}{\micro\ampere}. \autoref{fig:3}a displays the Hall voltage for the X, Y, Z1 and Z2 modes. As expected, the response is linear. The output voltage is in the microvolt order, as typical for Hall sensor devices in the millitesla range. In general, the sensitivity of Hall effect devices is reported normalized to the supply current or supply voltage. Two metrics known as current-related sensitivity ($S_I$) and voltage-related sensitivity ($S_V$) are reported in \autoref{fig:3}b as a function of the supply current. Considerable fluctuations, on the scale of 2 to \SI{5}{\volt\per\ampere\per\tesla}, are present because of the measurement noise. Therefore, related second order effects such as heating and JFET effect are not resolvable in this supply range. %The uncertainty ranges, indeed, do not intersect perfectly, which implies that some slight trend might be present. 

The sensitivity measurements were repeated for three different samples: P1, P2 and P3. The average current- and voltage-related sensitivities, as well as the input resistances measured at \SI{100}{\micro\ampere} for the three samples, are reported in \autoref{tab1}. The difference in sensitivity and resistance is due to process spread mainly caused by the non-uniform distribution of the spray-coated photoresist, which determines slightly larger or thinner features depending on the device position on the wafer.

\begin{table}
\centering
\caption{\textbf{Magnetic field sensitivity, crosstalk, and input resistance for two orthogonal phases of sensors P1 ... P4}}
\begin{tabular}{c|c|c|c|c c|c c c}
\multirow{2}{*}{Sensor} & \multirow{2}{*}{Mode} & {$S_I$} & {$S_V$}& \multicolumn{2}{c|}{$R$ (\SI{}{\kilo\ohm})} & \multicolumn{3}{c}{Crosstalk (\SI{}{\volt\per\ampere\per\tesla})}\\
 & & (\SI{}{\volt\per\ampere\per\tesla}) & (\SI{}{\milli\volt\per\volt\per\tesla}) & Ph1 & Ph3 & X & Y & Z\\
\hhline{=|=|=|=|= =|= = =}
\multirow{4}{*}{P1} & X & 81.7 & 17.1 & 3.74 & 5.74 & - & - & -\\
 & Y & 82.2 & 17.1 & 3.74 & 5.79 & - & - & -\\
 & Z1 & 111 & 21.4 & 4.71 & 5.74 & - & - & -\\
 & Z2 & 198 & 20.6 & 9.26 & 9.60 & - & - & -\\

\hline

\multirow{4}{*}{P2} & X & 73.7 & 15.9 & 3.58 & 5.58 & - & - & -\\
 & Y & 72.5 & 15.6 & 3.60 & 5.64 & - & - & -\\
 & Z1 & 103 & 20.0 & 4.57 & 5.62 & - & - & -\\
 & Z2 & 190 & 20.0 & 9.29 & 9.40 & - & - & -\\
\hline

\multirow{4}{*}{P3} & X & 64.1 & 14.8 & 3.30 & 5.23 & - & - & -\\
 & Y & 67.2 & 15.7 & 3.29 & 5.19 & - & - & -\\
 & Z1 & 94.8 & 19.6 & 4.28 & 5.30 & - & - & -\\
 & Z2 & 178 & 20.2 & 8.66 & 8.67 & - & - & -\\
\hline

 \multirow{4}{*}{P4\textsuperscript{*}} & X & 68.6 & 14.9 & 3.61 & 5.67 & - & 0.38$\pm$0.31 & 0.79$\pm$0.87\\
 & Y & 68.6 & 14.8 & 3.65 & 5.72 & -0.63$\pm$0.57 & - & 1.07$\pm$0.72\\
 & Z1 & 100.3 & 19.4 & 4.69 & 5.65 & 2.3$\pm$1.3 & 1.68$\pm$0.78 & -\\
 & Z2 & 190.4 & 19.7 & 9.68 & 9.64 & 3.3$\pm$2.7 & 3.8$\pm$1.7 & -\\

\end{tabular}
\flushleft
*$S_{I}$ extracted from \autoref{fig:3}c interpolation
\label{tab1}
\end{table}

The crosstalk of each mode with the remaining components of the magnetic field was evaluated by rotating a fourth sample (P4) in the magnetic field. Three rotation axis were employed, which coincide with the x-, y- and z-axis of the sample coordinate system. The rotation axis was aligned orthogonal to the applied field, and the output voltages of the modes were evaluated for each rotation angle and three magnetic fields: \SI{0}{\milli\tesla}, \SI{24}{\milli\tesla} and \SI{48}{\milli\tesla}. The Hall voltages (the output voltage minus the offset) for a magnetic field of \SI{48}{\milli\tesla} are reported in \autoref{fig:3}c. The rotation axes and the direction of the field are displayed in the inset of each graph. The curves of each mode were fitted to the function:

\begin{equation}
    V_{Hall} = (Csin(\phi + \psi) + Dcos(\phi + \psi))I_{supply}B
    \label{eq:interpolation}
\end{equation}

where $\phi$ is the rotation angle in radians, $\psi$ is the phase shift of the curve, and C and D are the current-related sensitivity or the crosstalk, depending on the mode. The results are reported in \autoref{tab1}. \autoref{tab1} also reports the average current-related sensitivity of each mode, extracted from different fits. The crosstalks vary from mode to mode and with the magnetic field component, but they are below \SI{3.7}{\percent} ($<$ \SI{5.5}{\volt\per\ampere\per\tesla}). The best crosstalk extracted is \SI{0.55}{\percent} (\SI{-0.38}{\volt\per\ampere\per\tesla}), and it was achieved with mode X with y-magnetic field applied.  

\begin{figure}[t!]
  \centering
  \includegraphics[width=\textwidth]{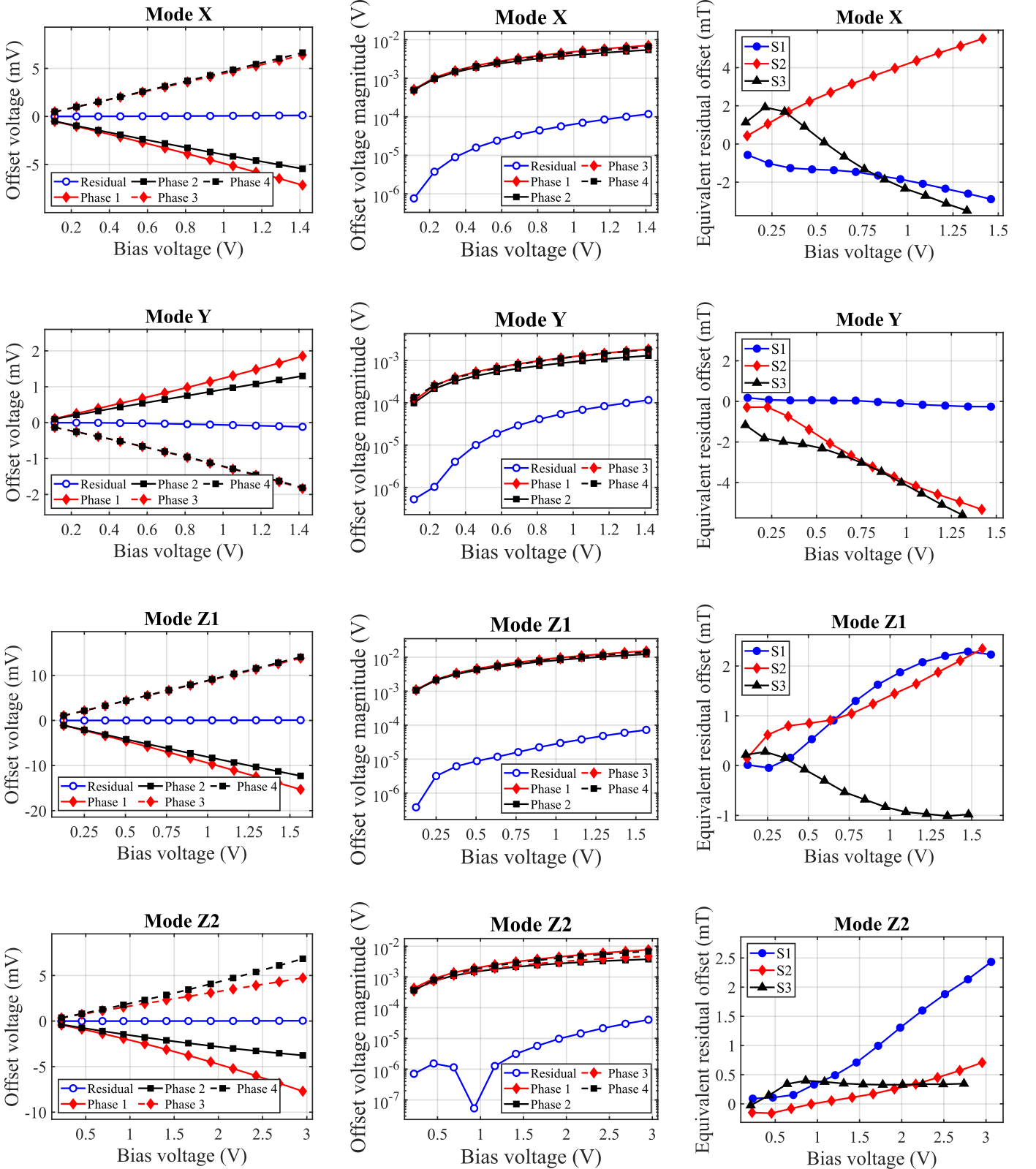}
  \caption{\textbf{Raw and residual offsets of the pyramid device.} The raw offsets of device P2 are reported in both linear and semilogarithmic scales, together with the residual offset. The equivalent magnetic residual offsets for the three sensors characterized are reported in linear scale.}
  \label{fig:4}
\end{figure}

\subsection*{Raw and Residual offset}

As previously mentioned, the tested devices display an offset voltage that can be effectively reduced by means of current spinning. To validate the spinning configurations, the output voltage is measured and averaged over the four phases of each mode's spinning sequence. In reality, the full spinning cycles include four additional phases to eliminate the offset of the multimeter, obtained by switching the sense contact polarity in each of the unique four phases. This adds redundancy to the spinning sequence which can be improved in microsystem implementation \cite{spinning_8phases}.
%The offset voltage is measured in a zero-Gauss chamber, shielded from electromagnetic interferences and external magnetic fields.
The measurement is performed for twelve equidistant currents in the range of \SIrange[]{25}{300}{\micro\ampere} and repeated for devices P1, P2, and P3. \autoref{fig:4} depicts the raw equivalent offset of the four phases, as well as the residual offset, on linear and logarithmic scales for modes X, Y Z1, and Z2. The raw offsets exhibit slight non-linear trends, which can be associated with second-order effects such as Seebeck voltage due to internal thermal distributions, the junction field effect, and packaging stresses.\cite{karen_res,udo_res} The four-phase spinning sequence effectively removes external thermal gradients and geometric offsets but is unable to remove the second-order terms, causing the residual offset. The residual offset achieved is stochastic and depends on the sample, mode, and supply voltage. For all characterized samples and modes, the offset reduction factor is between one and three orders of magnitude.

To coherently compare different devices and technologies, the residual offset is typically scaled by the sensitivity and reported in magnetic field units. This quantity, referred to as the equivalent magnetic offset, sets a bound on the device's limit of (zero magnetic field) detection. The equivalent residual offsets for all devices and modes are also represented in \autoref{fig:4}. No common or clear trend can be observed within the different samples for the same mode, apart from a general increase with the supply voltage. In addition, some curves do not even show a monotonic behavior. This suggests that the residual offset is not ruled by a single effect but rather by multiple physical causes that become more or less prominent at higher supply voltages. \cite{Bellekom}. At \SI{0.5}{\volt}, the general voltage supply at which VHDs are operated, the residual offset varies between \SIrange[]{0.1}{2.5}{mT}. 

\subsection*{Noise}

\begin{figure}[t!]
  \centering
  \includegraphics[width=\textwidth]{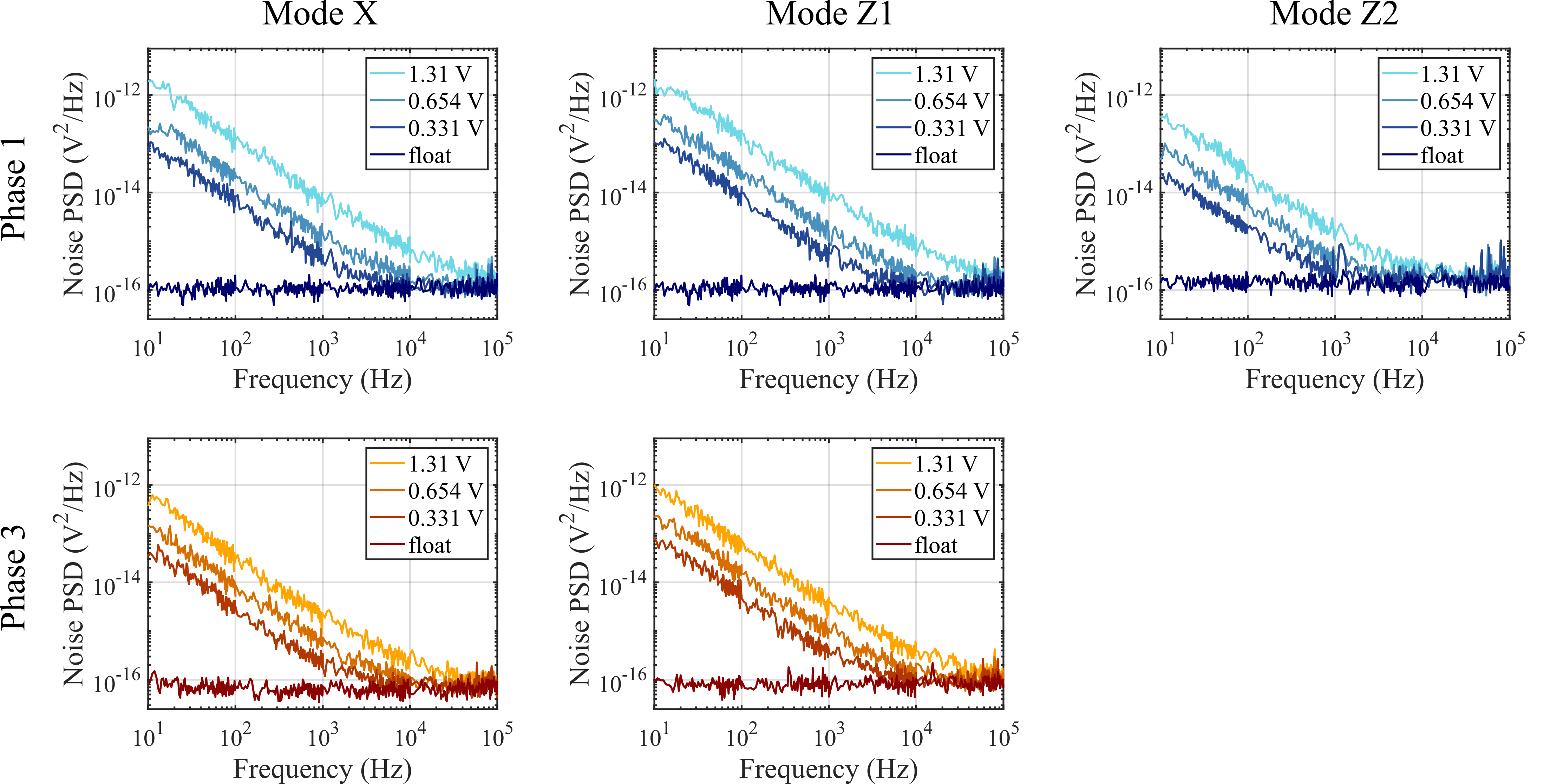}
  \caption{\textbf{ Noise power spectral densities (PSD) of sensor P1.} The noise is measured for three supply voltages at room temperature, in two orthogonal phases, for mode X, Z1 and Z2.}
  \label{fig:5}
\end{figure}

\begin{table}
\centering
\caption{\textbf{Power spectra fitted parameters at 1.31 V, extracted from P1.}}
\begin{tabular}{c|c|c|c|c|c}
\multirow{2}{*}{Mode} & \multirow{2}{*}{Phase} & \multirow{2}{*}{$\alpha_H/N$} & \multirow{2}{*}{$\gamma$} & Thermal floor & Corner frequency  \\
 & & & & (\SI{}{\nano\volt\per\sqrt{\hertz}}) & (\SI{}{\kilo\hertz})  \\
\hhline{=|=|=|=|= =}
\multirow{2}{*}{X} & Phase 1 & \SI{2.29e-11}{} & 1.25 & 10.4 & 27.4\\
 & Phase 3 & \SI{5.55e-12}{} & 1.23 & 8.07 & 16.0\\ 
\hline
\multirow{2}{*}{Z1} & Phase 1 & \SI{2.14e-11}{} & 1.21 & 10.4 & 38.0\\
 & Phase 3 & \SI{9.26e-12}{} & 1.22 & 9.02 & 22.1\\
\hline
\multirow{2}{*}{Z2} & \multirow{2}{*}{Phase 1} & \multirow{2}{*}{\SI{4.17e-12}{}}& \multirow{2}{*}{1.22} & \multirow{2}{*}{11.9} & \multirow{2}{*}{6.95}\\
 & &  &  & & \\
\hline

\end{tabular}
\label{tab3}
\end{table}

The final figure of merit evaluated for the novel Hall effect device is its noise performance. Alongside the offset, the noise defines the sensor's detection limit, but also its resolution. The noise spectral density was evaluated for modes X, Z1 and Z2, and at three different supply voltages: \SI{1.31}{\volt}, \SI{0.654}{\volt}, and \SI{0.331}{\volt}.  Mode Y is not reported since it is nominally identical to mode X. An additional measurement was taken with no supply voltage applied, with the input terminals of the pyramid device left floating. The frequency range for these measurements spanned from \SI{10}{\hertz} to \SI{100}{\kilo\hertz}.

For modes X and Z1, the noise spectral density was measured for two orthogonal phases, i.e. phase 1 and phase 3. Unlike mode Z2, these three modes exhibit asymmetric orthogonal phases, leading to distinct power density spectra that are intrinsically different. \autoref{fig:5} displays the characterization results.

As with many semiconductor devices, the sensor displays two primary noise components: flicker noise and thermal noise. Flicker noise can be described by the Hooge model, following the equation:
\begin{equation}
    S_{vf}^2 = V_{supply}^2 \cdot \frac{\alpha_H}{N} \cdot \frac{1}{f^\gamma},
\end{equation}
where $S_{vf}^2$ is the power spectral density of the flicker noise, $V_{supply}$ is the supply voltage, $\alpha_H$ is the Hooge parameter, $N$ is the total number of carriers, and $\gamma$ is a dimensionless factor that should be close to 1.
The intersection frequency between the extrapolated flicker noise and the thermal noise floor is known as the corner frequency. It has been demonstrated that current-spinning modulates flicker noise if the spinning frequency exceeds the corner frequency.
 $\gamma$ and $\alpha_H/N$ were extracted for a supply voltage of \SI{1.31}{\volt} by fitting the 1/f portion of the spectrum with the Hooge model. In addition, the corner frequencies and thermal floors were derived. These parameters are reported in \autoref{tab3}. For all modes and phases, the $\gamma$ factor is higher than 1, which implies the presence of traps with a nonuniform energy distribution \cite{dowling_noise}. The thermal noise differs between modes and phases, as expected with the varying input impedances reported in \autoref{tab1}, and it is around \SI{10}{\nano\volt\per\sqrt{\hertz}}. This corresponds to an equivalent magnetic noise level of around \SI{0.5}{\micro\tesla\per\sqrt{\hertz}} at \SI{1.31}{V}. The corner frequencies are below \SI{40}{\kilo\hertz}, for a supply voltage of \SI{1.31}{\volt}.

%Present the key findings of the study. Use figures and tables to illustrate the data where appropriate. Ensure all figures and tables are referenced in the text.

\section*{Discussion}

Based on the results obtained, we state that the device is a valid and promising solution for 3D magnetic field sensing. Although the proposed simplified model cannot provide an accurate quantitative description of the pyramid sensor, it highlights important similarities with planar Hall devices and provides a starting point for the sensor optimization. In fact, \autoref{current_sens} highlights how low doping is beneficial for both current- ($S_I$) and voltage-related ($S_V$) sensitivity. In addition, it hints that the performances are not degraded by shallow junction depths but rather enhanced. This is an evident advantage over vertical Hall devices, since VHDs require deep and narrow diffusions to operate, which are quite challenging to produce. On the other hand, the device relied on spin-coating for its manufacturing, which is not a process commonly employed in CMOS technology. Further technological development is therefore necessary to achieve CMOS integration.

A summary of the 3-axis Hall pyramid device performances, together with the existing state-of-the-art 3-axis sensors (depicted in \autoref{fig:1}) and 5-contact vertical Hall devices (5CVHD), is reported in \autoref{tab4}. 
The device exhibits high $S_I$, higher than several state-of-the-art devices such as the PHD + VHDs configuration and the hexagonal Hall device, but relatively low $S_V$. This can be attributed to the high doping of the device, approximately \SI{5e16}{\centi\meter^{-3}}, which degrades mobility and, thus, sensitivity. What is more, $S_V$ might be increased by improving the geometry of the device, i.e. choosing proper contact and pyramid sizes.\cite{Ruggeri} It is also worth noticing that the XY $S_I$ is lower than Z1, which in turn is lower than Z2 mode. In fact, in the first phase of mode XY, the current is split into two branches and both the x- and y-components of the field can be detected at the same time. This means that, for the same current, the sensitivity to each component is halved. Mode Z2, instead, displays a higher sensitivity because the current is pushed through one set of contacts instead of two, which inherently has higher input impedance.  The best $S_I$ and $S_V$ ratios between in-plane and out-of-plane modes are 0.74 and 0.83. Although the device is not optimized, it exhibits high isotropy which is already better than many existing 3-axis Hall sensors, such as the PHD+VHDs configuration and the IMC Hall. What is more, it can be further improved by a proper design. 
The crosstalks exhibited by the devices are reasonably low. As already explained, their origin is due to asymmetries in the structure. Despite sharing the same fundamental causes with the offset, cross-sensitivity cannot be current-spun. The only method to reduce this effect is to improve the manufacturing process, minimizing mismatches and inhomogeneities. 

These first-generation sensors, however, display very high offset and flicker noise. Within the probed voltage range, the raw offset is in the order of \SIrange{1}{10}{\milli\volt}, which corresponds to an equivalent field of around \SIrange{50}{100}{\milli\tesla}.
This suggests strong underlying asymmetries, which can be related to the uncommon process flow that involved spray-coated photoresist patterning and implantation of tilted (111) silicon planes. Despite this fact, the crosstalk remains below \SI{3.7}{\percent} of the sensitivity. This hints that, with proper process optimization, much lower crosstalk might be achieved. The residual offset ($B_{off,res}$) is in the millitesla order, which is similar to 5CVHDs.\cite{SANDER201692,SCHURIG200498,Ausserlechner} From a certain point of view the sensor, when used in X, Y or Z1 mode, can be seen as a "folded" vertical Hall device. It is also very interesting to notice that modes X,Y and Z1 present raw offsets that in phases 3 and 4 almost overlap, while in phases 1 and 2 they diverge already at low supply voltages. This recalls the behavior exhibited by vertical Hall devices \cite{PAUL201224}. On the other hand, mode Z2 exhibits more similarities with planar Hall devices than the vertical counterpart. For the same supply voltage, the residual offset is almost one order of magnitude better than the other three modes but still much higher than planar Hall devices \cite{vandermeer}. This suggests that some extra components of residual offset might be present in the sensor which are not shared with Hall plates. In addition, it is also worth noting that the device exhibits a relatively low active volume and footprint, compared to Hall plates and other state-of-the-art devices. Smaller devices tend to display larger raw offsets produced by lithographic inaccuracies, but also higher residual offsets due to velocity saturation caused by the higher electric fields. Therefore, the results obtained with large footprint devices, such as the hexagonal Hall sensor \cite{SANDER2016587}, and the pyramid device may not be directly comparable. A similar argument can be made for flicker noise, which scales down with the sensor's active volume. 
As already stated, the residual offset curves do not display a clear trend, so multiple sources are involved. However, some samples display a quadratic offset voltage increase (i.e. a linear equivalent offset increase) for high supply, which suggests Joule heating and strong temperature gradients. Other samples, however, reach a peak at a certain voltage and then steadily decrease, hinting at the competing mechanisms that push the offset to change polarity. The resistance non-linearity at \SI{0.5}{\volt} was $<\SI{4.7}{\percent\per\volt}$. This implies that the junction field effect is present and plays a role, but it is less prominent than for VHDs, which display non-linearities around \SI{10}{\percent\per\volt}.\cite{Ausserlechner}

The high flicker noise is not a significant problem since it can be modulated by the spinning sequence as long as the spinning frequency is higher than the corner frequency \cite{flicker_modulation}. 
The corner frequencies, \SIrange{6.95}{38.0}{\kilo\hertz} at \SI{1.31}{V}, are sufficiently low to properly spin the devices with interfacing electronics. However, the origin of the high flicker could highlight further problems and/or properties of the device. For instance, one of the possible causes of offset and flicker noise is mobile charges in the passivation layers, which move around in the presence of an electric field and behave as trap levels \cite{POPOVIC198939}. This is consistent with measured $\gamma > 1$, as well as with the high $\alpha_H/N$. In addition, the (111) sloped planes might present some extra trap levels due to the quality of the etched surface. In addition, asymmetric local heating and noisy contacts could be at play\cite{Barone2008-pn}, which also affect the residual offset. All these considerations and design guidelines are fundamental starting points for future work.

\begin{table}
\centering
\caption{\textbf{Comparison of the novel 3-axis pyramid device with the state-of-the-art}}
\begin{tabular}{c|c|c|c|c|c|c}
\multirow{2}{*}{Sensor} & \multicolumn{2}{c|}{{$S_I$ (\SI{}{\volt\per\ampere\per\tesla})}} & \multicolumn{2}{c|}{{$S_V$ (\SI{}{\milli\volt\per\volt\per\tesla})}} &$B_{off,res}$(\SI{1}{V}) & Spatial resolution\\
& XY & Z  & XY & Z & (\SI{}{\milli\tesla}) & (\SI{}{\micro\meter}) \\
\hhline{=|=|=|=|=|=|=}

Pyramid & 64.1-82.2& 94.8-198& 14.8-17.1& 19.6-21.4 & 0.2-4 & 50\\

\hline

5CVHD\cite{SANDER201692} & 40.5 & - & 17.2& -& 0.7-3.4 & -\\
\hline

5CVHD\cite{SCHURIG200498} & 46.4 & - & 32& -& 1-4 & -\\
\hline

5CVHD\cite{Ausserlechner} & $\sim120$ & - & 35.1& -& 0.6 & -\\
\hline

PHD+VHDs\cite{Pascal,SANDER2016587} & 6.27 & 90.1 & 1.9& 27.3& -& 44\\

\hline

3-axis planar\cite{SCHOTT2000167}& 46-827& 17-909&  20-54& 20-33& - & 50-500\\

\hline

Hexagonal\cite{Sander_table}& 8.6-8.8 & 8.7 & 33.0-33.9& 33.3 & 0.04 & 850\\

\hline

IMC Hall\cite{schott_IMC}& $\sim1500$ & 250 & $\sim300$& 50& $\sim0.03$\textsuperscript{*} & 2000\\
\end{tabular}
\label{tab4}
\flushleft\textsuperscript{*}Estimated from the data reported in the paper.\cite{schott_IMC}
\end{table}
\section*{Materials and methods}

\subsection*{Fabrication and Process flow}
\begin{figure}[t!]
  \centering
  \includegraphics[width=\textwidth]{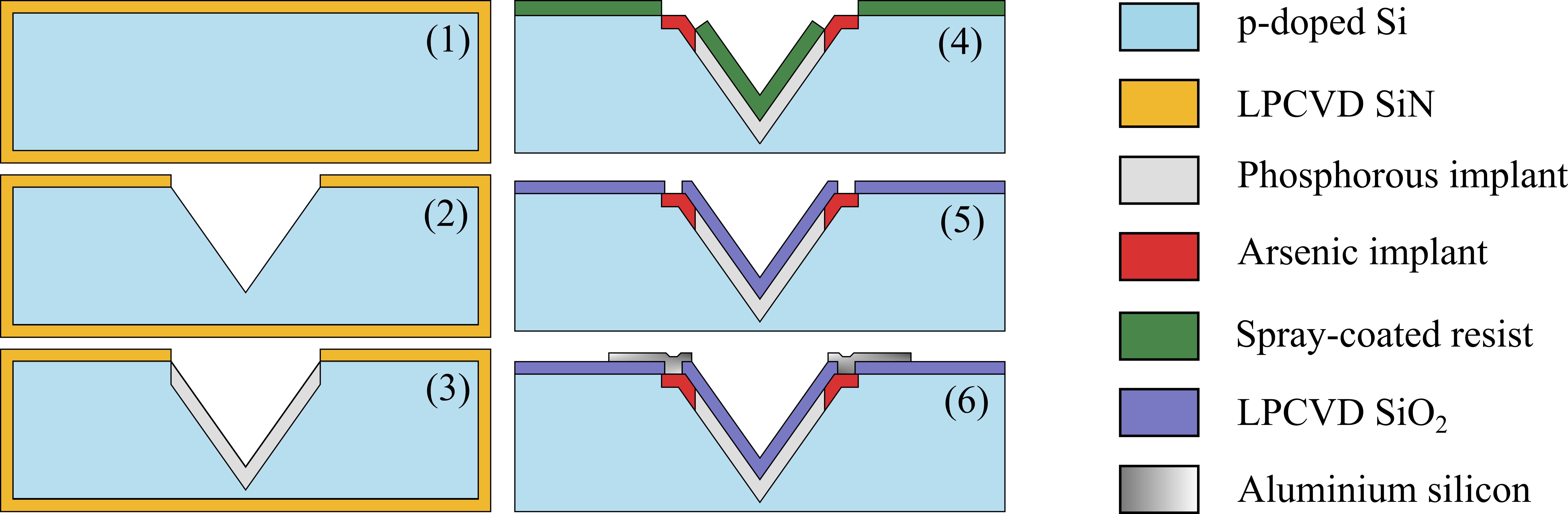}
  \caption{\textbf{Pyramid sensor process flow.} (1) LPCVD SiN deposition. (2) Pyramid cavity etching. (3) Active area implantation (4) n+ contact implantation (5) Device passivation (6) AlSi sputtering and patterning}
  \label{fig:6}
\end{figure}

The novel 3-axis sensors were realized with a process flow that leveraged CMOS technology and standard MEMS micromachining. The flowchart is displayed in \autoref{fig:6}. First, the p-doped (100) silicon wafer (resistivity from \SIrange[]{2}{5}{\ohm\cdot\centi\meter}) was covered with low-pressure chemical vapor deposited (LPCVD) silicon nitride (SiH\textsubscript{2}Cl\textsubscript{2} \SI{340}{sccm}; NH\textsubscript{3} \SI{60}{sccm}; \SI{850}{\degreeCelsius}). The nitride layer was patterned by reactive ion etching (RIE) using C\textsubscript{2}F\textsubscript{6}, followed by the anisotropic wet etching of the silicon with TMAH \SI{25}{\percent} at \SI{90}{\degreeCelsius}. A \SI{20}{\nano\meter} silicon oxide dirt barrier was then grown through dry oxidation at \SI{950}{\degreeCelsius}. After that, the pyramidal cavities were ion-implanted with phosphorus (energy: \SI{100}{\kilo\electronvolt};  dose: \SI{4.5e12}{\centi\meter^{-2}}, tilt: \SI{0}{\degree}), employing again the Si\textsubscript{3}N\textsubscript{4} layer as a self-aligned implantation mask. The nitride layer was removed with phosphoric acid at \SI{155}{\degreeCelsius}, and the devices were annealed at 
\SI{1150}{\degreeCelsius} for \SI{60}{\minute} to activate and diffuse the dopants. The goal net doping and junction depth were \SI{5e16}{\centi\meter^{-3}} and \SI{1}{\micro\meter}. After the growth of a second \SI{20}{nm} dirt barrier, the sloped sidewalls were covered with spray-coated photoresist and patterned to define the regions to be implanted with arsenic (energy: \SI{40}{\kilo\electronvolt};  dose: \SI{5e15}{\centi\meter^{-2}}, tilt: \SI{0}{\degree}). Spray-coated photoresist was employed to compensate for the inability of spin-coated photoresist to conformally cover the (111) sloped faces. The photoresist was removed, and 300 nm of LPCVD silicon oxide were deposited (TEOS bubbler \SI{50}{\degreeCelsius}; \SI{250}{\milli Torr}; \SI{700}{\degreeCelsius}) to play the role of passivation layer. The devices were annealed at \SI{1000}{\degreeCelsius} for \SI{30}{\minute} to repair the damage caused by implantation, activate the dopants and densify the oxide. The oxide was patterned with BHF 1:7 diluted with water to open the vias, and \SI{500}{\nano\meter} of  Aluminum-silicon (\SI{1}{\percent}) was sputtered onto the wafer at \SI{350}{\degreeCelsius}. The aluminum-silicon was wet-etched with PES (HNO\textsubscript{3}, H\textsubscript{3}PO\textsubscript{4}, acetic acid and water) to define the pads and the metal interconnections, and subjected to a polysilicon dip-etch with HNO\textsubscript{3} and HF to remove the polysilicon grains. Lastly, the metal lines were annealed at 400 °C in a reducing atmosphere (N2 3.0 L/min, H2 0.2 L/min). In the back end of line, the wafers were diced and the devices were glued and wirebonded to a carrier PCB.

\subsection*{Experimental setups and procedures}

Four experimental setups were employed in this work to measure sensitivity, crosstalk, offset, and power noise spectrum.

For the sensitivity measurements, a Helmholtz coil with a diameter of around \SI{30}{\centi\meter} was employed to apply a uniform magnetic field in the range \SIrange{0}{2}{\milli\tesla}. The coil was driven by a TENMA 72-2645 programmable power source. The sensor was supplied with current by a Keithley 2400 sourcemeter, and the output voltage was measured by an Agilent 
34401A multimeter. The spinning sequence was actuated by an Agilent U2751A switch matrix. The same power source, sourcemeter, multimeter and switch matrix were reused in the crosstalk and offset measurements. The measurements were repeated for twelve equidistant currents in the range \SIrange[]{25}{300}{\micro\ampere}, and for each current the eight-phase spinning sequence was repeated 20 times. The current-spun output was averaged over the 20 measurements performed. The alignment of the sample in the Helmholtz coil was performed through a 3D printed scaffold, placed and aligned by means of a FW Bell teslameter model 9500.
To perform the measurements for negative magnetic field, the supply terminals were swapped, to achieve the range \SIrange{-2}{0}{mT}. The offset measured during the sweeps is then subtracted from the output voltage, and the curves are interpolated with a linear regression to obtain the sensitivity. 
The supply voltage was averaged in the $S_V$ calculation.

For the crosstalk measurements, another Helmholtz coil with an inner diameter of around \SI{5}{\centi\meter} was used to reach a higher field up to \SI{48}{\milli\tesla}. The coil is driven with a current ranging from \SIrange[]{0}{2}{A}. The sample was carefully placed inside the coil and manually rotated by means of a 3D printed knob, with a graded scale, connected to the sample holder. The sample was rotated over 17 equidistant angles to perform a complete \SI{360}{\degree} rotation. The angular rotation error of \SI{3}{\degree} was extracted from the operator's resolving ability on the graded scale. Eventual misplacements in the initial \SI{0}{\degree} position are not significant  since every other rotation would be affected by the same systematic error and can be corrected through a phase shift in the interpolating function. For each rotation angle, the device was supplied with a \SI{100}{\micro\ampere}. The device was current-spun in each measurement, with a varying number of cycles that ranges from 2 to 21, depending on the mode and rotation angle. The output voltage was measured for three equidistant magnetic fields in the range \SIrange{0}{48}{\milli\tesla}, and the slope and intercept of the output voltage in function of the field were extracted. The offset was then removed from the output voltage to obtain the Hall signal. In this way, eventual magnetization effects, external constant magnetic fields, or drifts in the offset were effectively removed. The Hall signal in function of the rotation angle was then interpolated with \autoref{eq:interpolation} to obtain the sensitivity and crosstalk. 

For the offset measurement, the sample was placed in a zero-Gauss chamber to shield it from external magnetic fields. The current supply was swept from \SIrange{25}{300}{\micro\ampere} with step of \SI{25}{\micro\ampere}. For each current, every phase and current-spun output was averaged over 50 cycles.
The equivalent magnetic residual offset was obtained by dividing the residual offset by the sensitivity obtained from the Helmholtz coil measurements. The sensitivities used are the ones reported in \autoref{tab1}. The voltage reported on the x-axis of the graphs is the average supply voltage of the eight phases.

For the noise measurements, a HP4395A spectrum analyzer was used to extract the PSD. The device was supplied with \SI{1.3}{V} rechargeable battery. A voltage divider was adopted to split the voltage by half or one-fourth. The resistors used were approximately \SI{1600}{\ohm} and \SI{1600}{\ohm}, and \SI{2400}{\ohm} and \SI{800}{\ohm}. The voltage divider was low-pass filtered with a \SI{20}{\micro\farad} capacitor. The output voltage was then amplified by SR560 low-noise voltage amplifier. The battery-driven instrument was used in low-noise mode and AC coupling, with gain 200 and cutoff frequency \SI{1}{\mega\hertz}. To overcome the inability of the HP4395A spectrum analyzer to perform a sweep in logarithmic (frequency) scale, a list sweep was performed: for each decade, the noise power density was measured in 150 linearly spaced points. \SI{50}{\hertz} and integer multiples were skipped to avoid unwanted and unmeaningful peaks from the power grid. The sample and the biasing circuit were stored, during the measurement, in a zero-Gauss chamber to shield them against electromagnetic interferences and stray magnetic fields.
 The first 300 points, excluding the first 20, were fitted with the Hooge model to obtain the parameters reported in \autoref{tab3}. Only the \SI{1.31}{\volt} curves were fitted since they are the ones with the furthest corner frequency, so the 1/f fitting could be performed with a larger number of data samples.

%This section should describe the materials and methods used in the study in sufficient detail that others can replicate the work. Include descriptions of the experimental setup, procedures, and analytical techniques.

\section*{Conclusion}
In this work, we developed a novel type of 3-axis magnetic sensor based on the Hall effect and an inverted pyramidal geometry. The device was realized by combining MEMS micromachining and standard CMOS technology, and it exhibited high in-plane and out-of-plane current sensitivities in the range \SI{64.1}{} to \SI{198}{\volt\per\ampere\per\tesla} . The sensor presented low crosstalk, not higher than \SI{3.7}{\percent}. In addition, we proved that the device can be current-spun, achieving a reduction in raw offset of one to three orders of magnitude. The noise spectral density of the device was measured, and it exhibited corner frequencies lower than \SI{40}{kHz} at \SI{1.31}{V}, which means that the device can be spun with modern CMOS technology nodes without introducing extra errors due to the spinning circuitry. Despite the substantial reduction due to current spinning, the residual offset is still in the millitesla order, which is higher than standard Hall plates but comparable to vertical Hall devices. Significant improvement in the design and process flow is possible, but nevertheless the device is a promising solution and a simple alternative to state-of-the-art 3-axis Hall-effect magnetic sensors. The presented device, with proper development, will enable accurate and cost-effective position sensing and angle measurements for automotive, industrial and consumer electronics applications.

\section*{Acknowledgments}
This work was funded in part by the European Union. Views and opinions expressed are however those of the author(s) only and do not necessarily reflect those of the European Union or Marie Skłodowska-Curie Actions in the European Reserach Executive Agency. Neither
the European Union nor the granting authority can be held responsible for them. The authors gratefully acknowledge the Else Koi Lab staff for processing support, and the Electronic Instrumentation technicians for the help with the lab testing equipment. In addition, the authors would like to thank Leon Abelmann for the help with the magnetic field experimental setup, Jannik Strube and Floris van Mourik for useful discussions and assistance with the measurement setups.

\section*{Competing interests}
The authors declare no competing interests.

\section*{Author contributions}
JR designed, manufactured, and characterized the devices, with the help of KD in all steps. UA and HK provided guidance with the measurements and results interpretation. KD supervised the project. JR drafted the manuscript and received feedback from all authors.

\bibliography{biblio} % Your bibliography file

% Example of manually entered references
% \begin{thebibliography}{99}
% \bibitem{ref1} Author1, A., Author2, B. Title of the paper. Journal Name, Year, Volume, Pages.
% \bibitem{ref2} Author1, A., Author2, B. Title of the paper. Journal Name, Year, Volume, Pages.
% \end{thebibliography}

\end{document}